\documentclass[%
 reprint,
superscriptaddress,
 amsmath,amssymb,
 aps,prper
]{revtex4-2}

\usepackage{graphicx}% Include figure files
\usepackage{dcolumn}% Align table columns on decimal point
\usepackage{bm}% bold math
\usepackage{geometry}
\usepackage{graphicx}
\usepackage{float}
\usepackage{amsmath}
\usepackage{amssymb}
\usepackage{array}
\usepackage{subfiles} 
\usepackage{outlines}
\usepackage{multirow}
\usepackage{longtable}
\usepackage{textgreek}
\usepackage[table]{xcolor}
\usepackage{booktabs}

\begin{document}

\preprint{APS/123-QED}

\title{Empirical evidence of the inseparability of mathematics and physics in  expert reasoning about novel graphing tasks}% Force line breaks with \\
%\thanks{A footnote to the article title}%

\author{Charlotte Zimmerman}
\affiliation{Department of Physics, University of Washington}%
\author{Alexis Olsho}
\affiliation{Department of Physics and Meteorology, US Air Force Academy}%
\author{Michael Loverude}
\affiliation{Department of Physics, California State University Fullerton}%

\author{Suzanne White Brahmia}
\affiliation{Department of Physics, University of Washington}%

\date{\today}% It is always \today, today,
             %  but any date may be explicitly specified

\begin{abstract}

%Pre-college mathematics modeling instruction often frames mathematical reasoning as a separate process from reasoning about the real world---commonly framed as separate stages of a reasoning cycle. However, evidence in physics and mathematics education research has shown that this separation doesn't represent how upper-division physics students successfully model; there is much more interdependence of these worlds in their sense-making. It is likely that physics instructors---those designing introductory physics courses---also typically reason in a more blended way. This difference in approach may lead to a mismatch in expectations, and frustration, for both students and instructors. In order to help characterize physics experts' mathematical reasoning, we examine how experts use mathematics in physics contexts while developing graphical models and report on it in this paper. An important finding is that there was essentially no evidence of experts reasoning in a context-free way with these tasks, but instead they used physical reasoning---either grounded in the context of the task or from abstract physical models---to guide their mathematics. We contribute to the body of knowledge about how mathematical reasoning appears in physics. We aim to help researchers and instructors recognize the connections to the physical world in expert mathematical reasoning, and thereby be better equipped to develop materials and methods that can help students start building those connections as well.

Pre-college mathematics modeling instruction often frames mathematics as being separated from reasoning about the real world---and commonly treats reasoning mathematically and reasoning about the real-world context as separate stages of a modeling cycle. In this paper, we present evidence that helps characterize how experts use mathematics in physics contexts while developing graphical models. An important finding is that there was essentially no evidence of experts reasoning in a context-free way with these tasks, but instead they used physical reasoning---either grounded in the context of the task or from abstract physical models---to guide their mathematics. The difference in approach of physics instructors and students may lead to a mismatch in expectations, and frustration, for both parties. This work contributes to the body of knowledge about how mathematical reasoning appears in physics, and can help researchers and instructors recognize the connections to the physical world in expert mathematical reasoning. This can help instructors and researchers be better equipped to develop materials and methods that can help students start building those connections as well.

\end{abstract}

\maketitle
\section{Introduction}
\label{sec:intro}
Instruction about modeling in pre-college mathematics courses characteristically represents  ``mathematical'' reasoning as distinct from reasoning about the real world \cite{Usiskin2011MathematicalCurriculum, Usiskin2015MathematicalMathematics, Martin2016MathematicalSeries, Greefrath2022MathematicalTeaching}. Emphasizing a separation between mathematical reasoning and reasoning about the context during instruction may leave future physics students with the impression that there is physics-less mathematics associated with solving introductory physics problems. Recent research reveals that physics students \textit{perceive} that there is a difference between the mathematics they do in mathematics courses, and working with mathematical models in physics \cite{Bajracharya2023StudentPhysics, Taylor2023IMath}.

There is growing evidence that upper-division physics and engineering students do not demonstrate this separation in their practice when they are successful in their problem solving. There is, in fact, a continuous interplay of mathematical and physical reasoning \cite{Ferri2006TheoreticalProcess, Arleback2009OnSchool, Czocher2016IntroducingThinking, Serbin2022TheProblems}. We suggest that the physics experts who design and teach introductory physics courses may also 
reason in a less binary way, with more nuance in the ways that they demonstrate how to develop and make sense of models---and they may expect that their students are prepared to follow along. If instructors are modeling in physics in ways that are inconsistent with students' pre-requisite mathematics courses, both students and instructors may be frustrated with the learning outcomes. Therefore, 
understanding the interface between expert modeling in introductory physics and students' preparation from prior mathematics courses can help improve the learning and teaching of physics for both students and instructors. To this end, this work sets out to better understand expert modeling in the contexts of challenging, introductory-level, graphical tasks.  

Initially, we intended to isolate and characterize expert behavior when reasoning ``purely mathematically'' in introductory classical physics contexts. It was quickly evident that when experts modeled in the contexts of these tasks,  physical reasoning was ubiquitous and inseparable from mathematical reasoning. This finding led to a shift of our research focus from examining when physics experts engaged in ``purely mathematical reasoning'' to characterizing what they actually do. We sought to answer a research question that represents the core of this paper: ``What are some features of the mathematical reasoning used by physics experts when modeling graphically?'' 

The results we present contribute to the growing body of knowledge about how physics reasoning interacts with mathematical reasoning. Examining the reasoning of physics experts provides researchers with data from one end of the expert-novice continuum, and may help physics instructors recognize their own reasoning patterns, assumptions of comprehension, and emergent expertise in their own students.

\section{Background}
\label{sec:background}
There is a growing consensus that quantitative reasoning in pure mathematics contexts is different from that in physics contexts \cite{Sherin2001HowEquations, Kuo2013HowProblems, Hu2013, WhiteBrahmia2014, Redish2015LanguageEpistemology, Caballero2015UnpackingHere, Bajracharya2016AnalyticalProblems, VanDenEynde2019, Serbin2022TheProblems}. Physics education researchers have proposed multiple characterizations of this difference: Redish described physics as a linguistic dialect distinct from the language of mathematics \cite{Redish2015LanguageEpistemology}; Caballero pointed to how the operationalization of mathematical tools is often different in physics than in mathematics \cite{Caballero2015UnpackingHere}; and White Brahmia emphasized the centrality of \textit{quantification} to mathematical modeling in physics contexts, 
where quantification refers to characterizing a feature of a system with a quantity---represented by a letter (e.g., $x,v,a$)---that includes a unit of measure and typically a sign.
\cite{Thompson2011a, Czocher2021AttendingSpace, WhiteBrahmia2017SignedMeaning,WhiteBrahmia2019QuantificationPhysics}.

Connecting a quantity's physical meaning to its mathematical representation is an important part of developing and making sense of mathematical models, and has been shown to be challenging for many students in introductory physics contexts and beyond \cite{Torigoe2011, Olsho2019TheSign, WhiteBrahmia2019QuantificationPhysics, WhiteBrahmia2021PhysicsPhysics, Czocher2021AttendingSpace}. Research has demonstrated student difficulties with making sense of the physical meaning of vectors, noting that many students can make sense of the symbols but struggle with translating between vector and graphical notation and extrapolating the information to a vector field \cite{BollenStudentFields}.
In calculus-based graphical contexts, Bajracharya and Thompson describe how successful students were able to connect the ideas of area under the curve and integration, demonstrating both procedural and conceptual understanding of the underlying mathematics, while also relating the mathematics to its physical meaning \cite{Bajracharya2016AnalyticalProblems}. These findings are aligned with research in physics education that facility with multiple mathematical representations is a key element of making sense of the meaning of mathematical models \cite{Hestenes1997ModelingTeachers, VanHeuvelen2001MultipleProcesses, Kohl2006EffectSkills, Rosengrant2006AnRepresentations, Brewe2008ModelingPhysics}.

\begin{figure}
    \centering
    \includegraphics[width=0.45\textwidth]{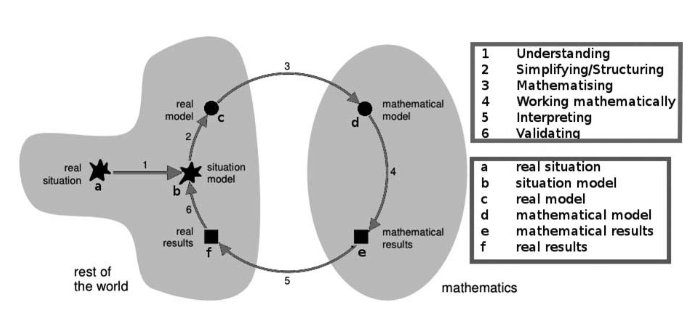}
    \caption{Czocher's redraft of Blum and Lei\ss's modeling cycle \cite{Blum20075.1Problems, Czocher2016IntroducingThinking}.}
    \label{fig:mathcycle}
\end{figure}

A common way to represent the sequence of moves in modeling is as a ``modeling cycle'' in which quantification is an early step, and \textit{validation}---interpreting a mathematical model within the physical context---is the last (see the example from Blum and Lei\ss~in Figure~\ref{fig:mathcycle}). Notably, the mathematics space depicted is accessed between quantification and validation, and does not involve consideration of the physical context, but instead represents mathematical manipulation in absence of attention to physical meaning. Similar cycles that separate mathematical reasoning and reasoning about the real world have been proposed both in physics and in mathematics education research \cite{Blum20075.1Problems, Redish2005ProblemCourses, Redish2008LookingEngineers, Uhden2012ModellingEducation, Wilcox2013AnalyticPhysics, Martin2016MathematicalSeries}. While recent education research has moved toward a more blended perspective of mathematics and physics reasoning, these cyclic representations commonly frame modeling education in high school mathematics courses and thereby influence students' perceptions of mathematical modeling before coming to college \cite{Usiskin2011MathematicalCurriculum, Greefrath2022MathematicalTeaching}.

Research has found little evidence of a clear separation between reasoning mathematically and reasoning about the real world in quantitative modeling; moreover, the progression of actual modeling may be messier than the published cycles would suggest \cite{Ferri2006TheoreticalProcess, Arleback2009OnSchool}. In a recent study, Czocher conducted interviews throughout an academic term of four engineering majors enrolled in a differential equations course \cite{Czocher2016IntroducingThinking}. In each interview, the students were observed solving problems in physics contexts that required generating mathematical descriptions from a variety of branches of mathematics, including differential equations. The author described a much finer-grained blending of mathematical reasoning and physical sense-making than is represented in \textit{a priori} ``modeling cycles.'' Specifically, Czocher notes that ``there are transitions that appear out-of-order. This was largely because three of the modeling transitions (understanding, simplifying/structuring, and validating) appeared early and often throughout the students’ modeling processes'' \cite{Czocher2016IntroducingThinking}.
Serbin and Wawro examined junior and senior physics students' reasoning in quantum mechanics through the lens of Uhden et al.'s modeling cycle, and similarly found that successful modeling patterns did not support separating mathematical reasoning and physics reasoning as predicted in the theories of how students generate models. The authors state ``when students draw on their mathematical knowledge\ldots it relies on their understanding of how the mathematical concepts and physics concepts are intertwined'' \cite{Serbin2022TheProblems}. 

The results of Czocher, and Serbin and Wawro suggest that modeling cycles that maintain separate mathematics reasoning and physics reasoning may require revision to reflect a tighter blend between ideas that are physics-like and math-like \cite{Czocher2016IntroducingThinking, Serbin2022TheProblems}. Typically, conceptual blending is used in physics education research to describe a fine-grained approach where particular utterances, equations, or terms are separated into mathematics or physics ``input spaces'' by the researcher \cite{Fauconnier2003ConceptualMeaning, Bing2007, Hu2013, Huynh2018BlendingNegative-ness,  Taylor2018SoKinematics, Schermerhorn2018InvestigatingSystems, VanDenEynde2020, Odden2021HowPhysics}. These input mental spaces are described as subconsciously blended into one idea by the speaker. The blended space may also include utterances that are inseparable by the researchers. Increasingly, physics education researchers are using the term more broadly to refer to the blended nature of mathematical structure and physical meaning \cite{WhiteBrahmia2021PhysicsPhysics, Olsho2019TheSign, Eichenlaub2019, Loverude2018MathematizationCourse}. 

There remains, however, a question in physics about the prevalence of blended reasoning as compared to ``purely mathematical'' reasoning---i.e. mathematics free of contextual meaning---and what moving between these reasoning modes might look like for physics experts.  The research we present here focuses on expert modeling in the contexts of graphing tasks, adding to current findings associated with the blended mental space between physics and mathematics in modeling. 

\section{Methods}
\label{sec:methods}
In this section, we describe the methods used to conduct and analyze the think-aloud interviews that were part of our investigation into how expert physicists reason quantitatively during graphing tasks \cite{Zimmerman2023ExpertTasks}.

\subsection{Data Collection}
We report on results from 15 think-aloud, individual interviews that were conducted with 10 graduate students and five faculty members at an R1 university in the Pacific Northwest. Of the 15 experts, six were female or non-binary identifying and nine were male identifying. The pool of experts also spanned a variety of subfields, and included both experimentalists and theorists. An effort was made to include diversity across race and country of origin. Participants were solicited by email, and offered a small gift card to a local coffee shop as a thank you for their time and effort.  

The graduate student interview transcripts were divided into two groups of five for data analysis. The first included early career graduate students who were enrolled in required graduate courses; the second, late career graduate students who were no longer taking courses outside their research area and were fully engaged in research.

\begin{figure}
    \centering
    \includegraphics[width=0.45\textwidth]{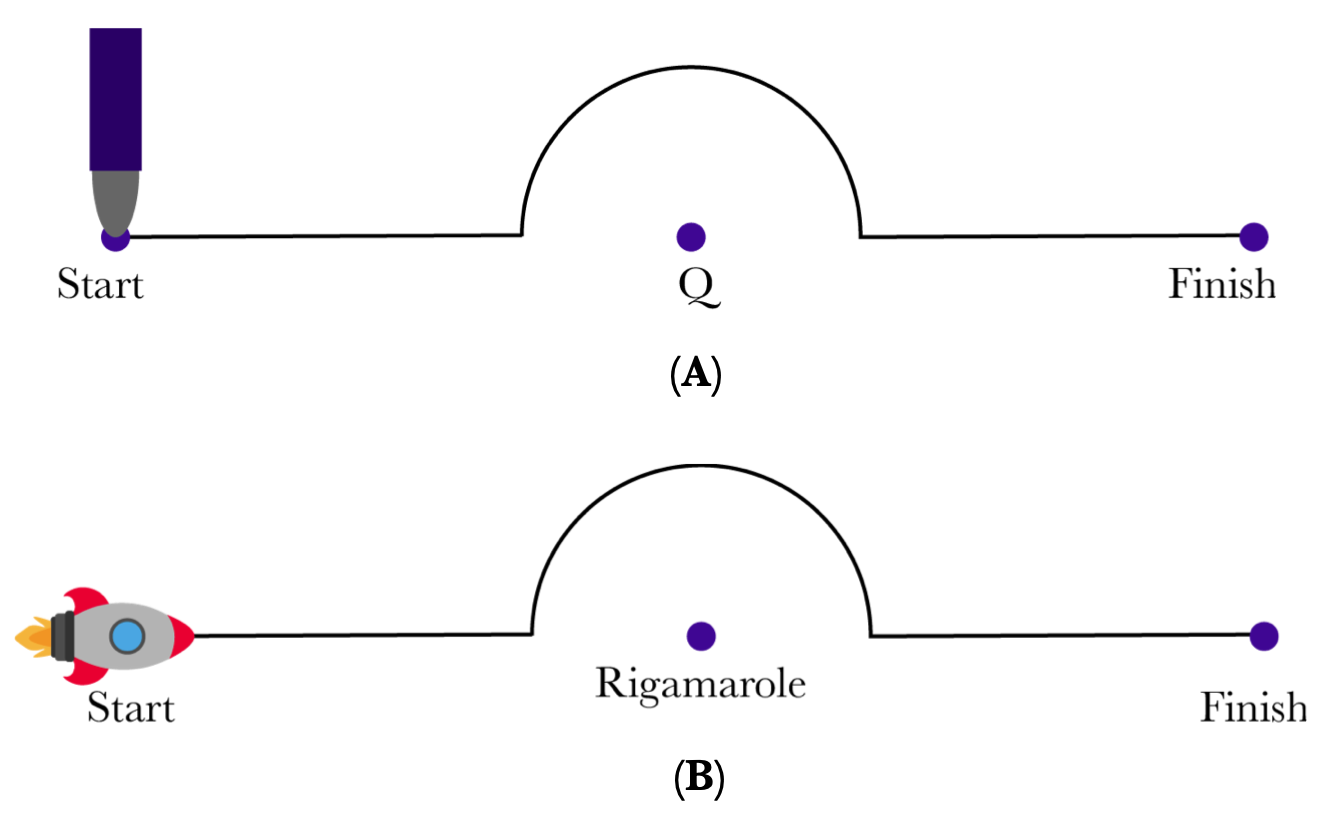}
    \caption{Stills from the animations associated with (A) Electric Charge and (B) Gravitation. The tasks prompt participants to create a graph that relates either the electric or gravitational potential and the total distance traveled of the probe or spaceship, as it moves at constant speed from start to finish.}
    \label{fig:potentialtasks}
\end{figure}

\begin{figure}
    \centering
    \includegraphics[width=0.45\textwidth]{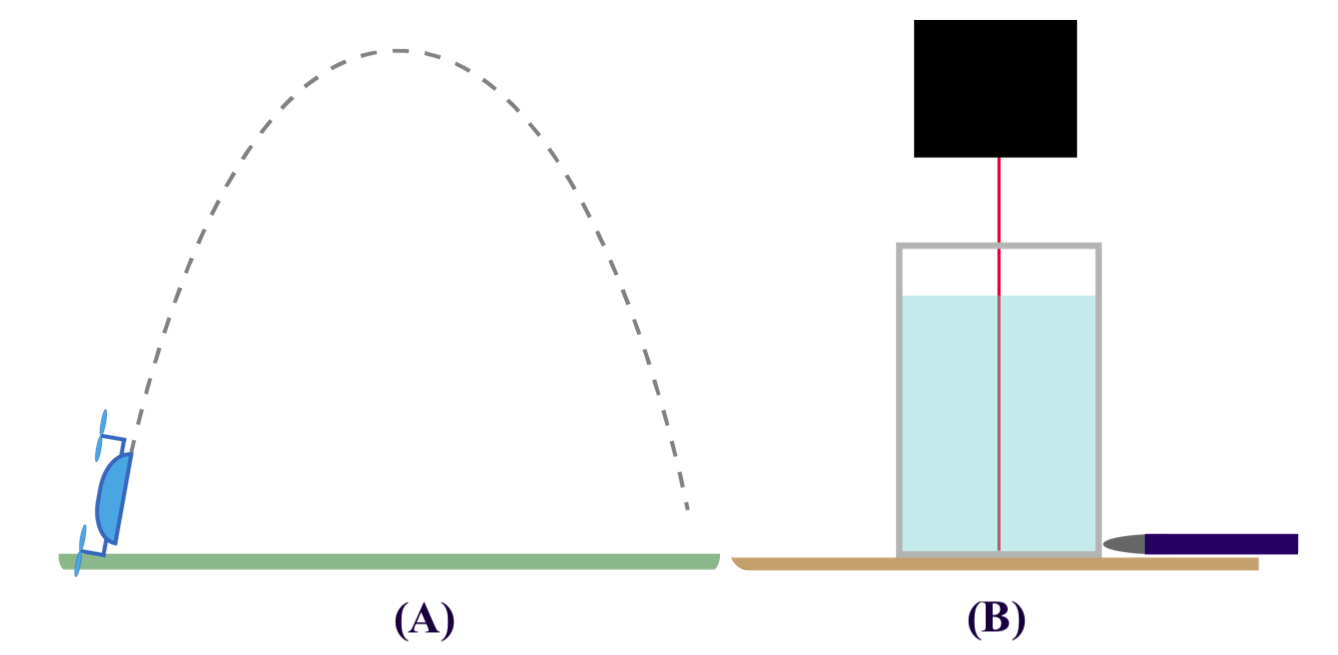}
    \caption{Stills from the animations associated with (A) Drone, in which participants were prompted to create a graph that relates the height to the angle of the drone, flying in the arc shown, and (B) Intensity, in which participants were prompted to create a graph that relates the intensity measured by the probe at the side of the liquid column to the total distance traveled of the probe as it moves from the bottom of the column to the top.}
    \label{fig:droneandlight}
\end{figure}

One member of the research team conducted the interviews, during which the participants were asked to solve 3--4 graphing tasks. In the analysis described in this paper, we focus on three tasks that were identical (or isomorphic) across the 15-person expert pool. For the first task, experts were given one of two isomorphic versions: ``Electric Charge'' or ``Gravitation'' (Fig.~\ref{fig:potentialtasks}). For the second and third tasks, all experts were given both ``Drone'' and ``Intensity'' (Fig.~\ref{fig:droneandlight}). All of the tasks prompted the participant to watch an animation and create a graph that related two quantities in the animation. 

Each interview lasted between 25 and 50 minutes, and ended when the participant expressed that they were finished with all the tasks. At the end of the interview, the interviewer asked clarifying questions, if applicable. The interviews were audio recorded; initial transcripts were created automatically using the Otter.ai software program \cite{2019Otter}, and were subsequently hand-corrected.

\subsection{Data Analysis}

\renewcommand{\arraystretch}{1.5}
\renewcommand{\tabcolsep}{8pt}
\begin{table*}[]
    \centering
    \begin{tabular}{p{0.3cm} | p{2cm} p{5.3cm} p{5.3cm}}
        \toprule
        & Code & Description & Example \\
        \cline{2-4}
        & Symbolizing
                & Representing quantity with symbols.
                & ``This is potential, $V$.''\\
        \multirow{3}{*}{\rotatebox[origin=c]{90}{\hspace{-1cm} QUANTITY}}
        & Mathematical Structure
                & Choosing a particular structure or discussing why a structure makes sense for a particular quantity.
                & ``So this is only going to be positive.''\\
        & Composite Quantity
                & Defining a quantity by others, or making sense of why a quantity is constructed of others.
                & ``$I$ is gonna be the number sent in times $e$ to the minus $y$ over $y_0$, where $y_0$ is a characteristic, sort of, thickness of the material.''\\
        & Proxy Quantity
                & Using one quantity in place of another to make sense of or visualize what is happening.
                & ``So as I go from distance traveled, I'm basically starting at a fixed $R_0$ and I'm going to a smaller $R$ [from] there.''\\
        \hline
        & Code & Description & Example \\
        \cline{2-4}
        \multirow{2}{*}{\rotatebox[origin=c]{90}{TASK CONTEXT}}
        && \\
        & Connection to the Task
                & Making sense of a quantity or feature of their graph, or drawing a diagram to visualize a quantity in the context of the task.
                & ``And so when they're far apart, there's low gravitational potential energy.'' \\
        && \\
        \hline
        & Code & Description & Example \\
        \cline{2-4}
        \multirow{3}{*}{\rotatebox[origin=c]{90}{MODELS}}
        & Compiled Model
                & When participants bring in a previously known model from physics.
                & ``Okay, so I'm going to sort of invoke Bier’s law.'' \\
        & Compiled Trend
                & When participants refer to a particular function and/or discuss its behavior.
                & ``I'm going to say this is a graph of something like a minus arctan of $x$.'' \\
        \bottomrule
    \end{tabular}
    \caption{Codes associated with the code categories Quantity, Task Context, and Models.}
    \label{tab:codes}
\end{table*}
The data were coded using a thematic analysis framework \cite{Nowell2017ThematicCriteria}. Thematic analysis emphasizes consistent communication across the research team to ensure reliability while recognizing that the research team may bring perspectives to the work based on prior experience with current literature. The practice assumes no \textit{a priori} coding scheme and focuses on grounding the codes in the data through iterative review and discussion. Our interpretation of these data is informed by our prior experiences, established theories of mathematical reasoning \cite{Blum20075.1Problems, Redish2005ProblemCourses, Redish2008LookingEngineers, Uhden2012ModellingEducation, Wilcox2013AnalyticPhysics}, mathematics education research testing the cyclic nature of these theories \cite{Czocher2016IntroducingThinking, Serbin2022TheProblems}, and research in physics education that uses conceptual blending to describe mathematical reasoning in physics contexts \cite{Bing2007, Hu2013, Huynh2018BlendingNegative-ness, Taylor2018SoKinematics, Loverude2018MathematizationCourse, VanDenEynde2019, Olsho2019TheSign, Eichenlaub2019, WhiteBrahmia2021PhysicsPhysics} as discussed in Section~\ref{sec:background}. 

One member of the research team coded the data, using participants' written work and notes from the interviewer to confirm the interpretation of the audio transcripts. Between coding cycles, the researcher conferred with the research team to establish wide-spread agreement. 

Once the codebook was stable, the same iterative process was followed to seek themes amongst the codes. Several code categories emerged, three of which we share here:
\begin{itemize}
    \item[(1)] \textbf{Quantity}, which refers to when participants generate, relate, symbolize, or discuss the mathematical structure of quantities which are inherently situated in the real world.
    \item[(2)] \textbf{Task Context}, which refers to when participants made explicit reference to the physical context of the task.
    \item[(3)] \textbf{Models}, which refers to the ways in which participants use previously known physical models or apply their knowledge of particular functions common in physics.
\end{itemize}
Table~\ref{tab:codes} shows the codes in these categories, along with a short description and example of each code. 

The research team took care to allow the codes to emerge from the data. Upon first glance, a code for ``physical reasoning'' seemed appropriate, however it quickly became clear that the ``physical reasoning'' code was ubiquitous across the data, rendering it meaningless. The lead coder also felt they could not distinguish between ``physical'' and ``mathematical'' reasoning throughout the data, even after lengthy discussion with the research team. Therefore, we instead developed the code Connection to the Task, and established the boundary that this code only included instances in which experts explicitly relate their reasoning to the task at hand. This distinction removed some of the subjective interpretation associated with deciding what kind of reasoning is physical, and grounded the coding in the statements of the experts. A consequence of this decision is that reasoning about an abstract physical model is not included in the Task Context category. For example, one expert stated:
\begin{quote}
    ``If I think about $v_x$ over $v_y$, $v_x$ is a constant and $v_y$ is linear in $t$.''
\end{quote}
This statement is categorized as Quantity, since the expert is symbolizing velocity quantities, but not Task Context since they are reasoning using the kinematic equations in abstract terms. Instead, we applied a separate code, Compiled Models, to indicate they were using a previously-known model \cite{Zimmerman2023ExpertTasks}. In contrast, later in their reasoning this expert stated:
\begin{quote}
    ``As we get towards the top...as my height increases, my angle---does it change faster?''
\end{quote}
It is clear the expert is considering the context of the task at hand when they talk about ``the top'' and the statement is categorized as Task Context.

\begin{figure*}
    \centering
    \includegraphics[width=\textwidth]{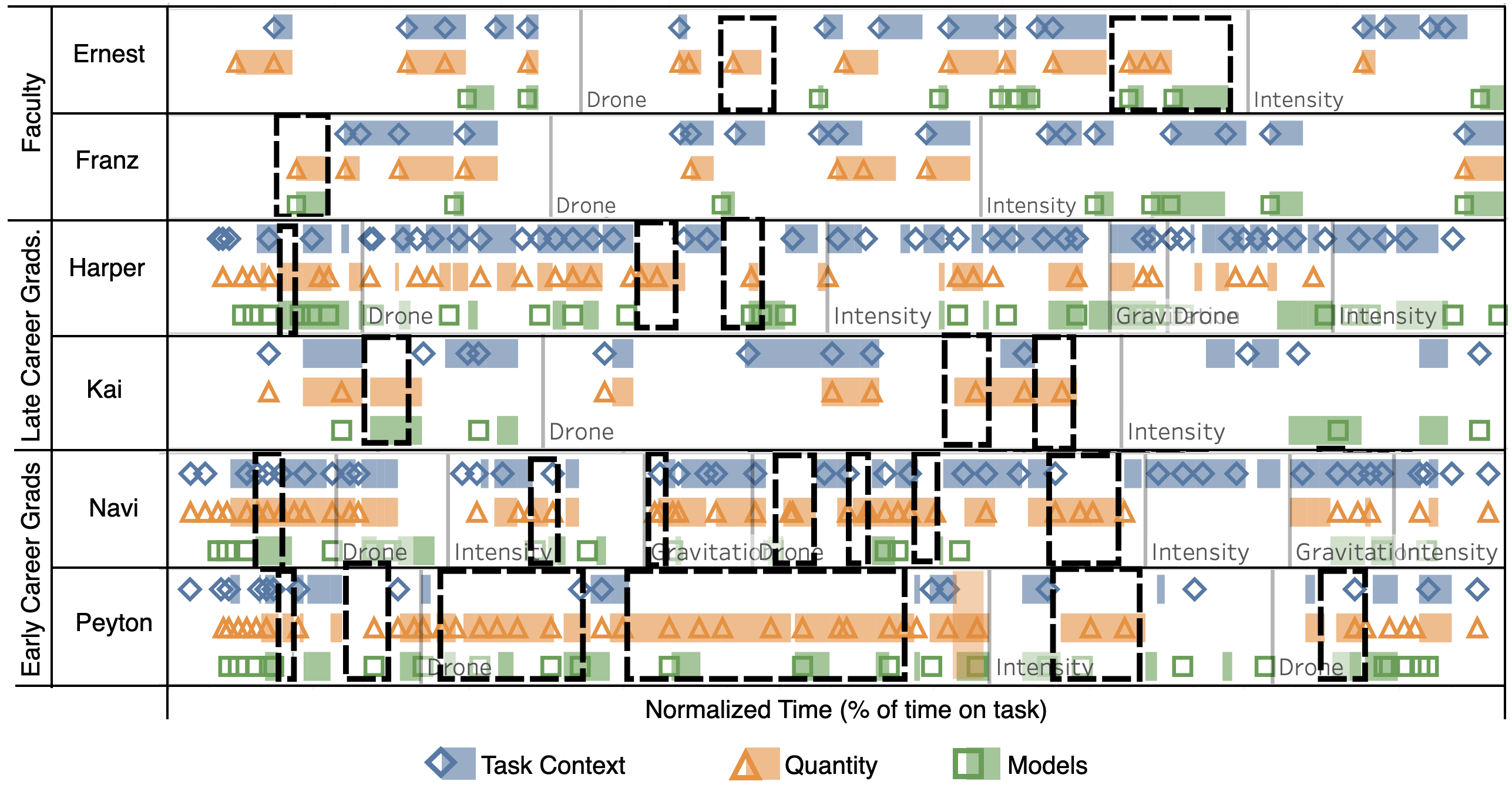}
    \caption{A visual representation of when participant statements were assigned the code categories Quantity, Task Context, and Models. The length of the bars corresponds to the time the participant was engaged in that line of reasoning. The horizontal axis represents time, normalized to the length of each interview. Therefore, it can be thought of as a percentage of their total time on task. The highlighted regions are where Quantity appears without Task Context. Vertical lines represent when participants switched tasks.}
    \label{fig:ganttCharts}
\end{figure*}

We created timeline charts to seek patterns between code categories. The timeline charts visually represented segments of transcript for when a participant's reasoning was assigned a particular code category, and the duration for which they were coded in that category. Figure~\ref{fig:ganttCharts} shows six representative timelines from our data. We sought patterns in which categories frequently appeared together and frequently appeared apart. 
After identifying segments of transcript in which a given pattern appeared (e.g., context coded with quantity), we returned to the transcripts for further analysis. We categorized the transcript segments into two groups: (1) those in which the text supported the pattern and (2) those in which the pattern was refuted. For example, for a pattern that appears together in the timelines, we collected all transcript pieces where the code categories coincided, and separately collected all transcript pieces where the code categories had not. We examined these sections of text carefully to find patterns in the ways that experts reasoned in order to characterize, or dismiss, the relationship identified in the timeline.

Finally, we used quantitative measures to compare the frequency of categories appearing separately or together. For example, we calculated the time on task for all utterances where Quantity had been coded (on its own, or with any other category), and the time on task where Quantity and Task Context had been coded together.

\section{Findings}
\label{sec:results}

Our research question for this study focuses on the relationship between physics and mathematics reasoning: ``What are some features of the mathematical reasoning used by physics experts when modeling graphically?''
Our data revealed that there was almost no time during which physics experts reasoned without reference to the physical world. This is in contrast to what is suggested by the presence of ``purely mathematical reasoning'' in common modeling cycles. Instead, the distinction between how experts reasoned physically and mathematically lay in whether they reasoned with reference to the physical world abstractly or with reference explicitly to the context of task at hand. 

We identify two major features of mathematical reasoning in physics from this work. (1) When physics experts reasoned mathematically, they generally did so with explicitly reference to the context of the task. (2) During the relatively infrequent occasions when experts did not attend explicitly to the task context, physics experts were likely to use other familiar physics models as a starting point to find an approximate, analytical solution. 

In the time segments during which physics experts did not consider the task context, we observed a behavior that we have named ``mathematical riffing'': experts were seen to algebraically manipulate a model to see if it could be adjusted to fit the novel context of the task. We found that  early career graduate students spent a larger percent of time-on-task riffing, as they were less likely to reject an unproductive model and thus spent longer trying to make the unproductive model work.  In this section, we describe each of these findings in detail. 

\subsection{Blended Mathematics and Physics}
A dynamic interplay between behaviors categorized by Quantity codes and Task Context codes emerged across all experts. The timeline charts (Figure~\ref{fig:ganttCharts}) indicate that behaviors coded as Quantity and Task Context appeared together frequently. The Quantity and Task Context categories were assigned at the same time when the research team could not separate the ideas within a single utterance. For example, one expert working on Gravitation stated:
\begin{quote}
``It's got to be negative one over R, because I want it to fall in.''
\end{quote}
\noindent
The speaker associated their symbolic expression with what it implies is happening in the physical system---the negative sign is required for the model to describe an attractive gravitational field that the spaceship can ``fall in'' to.

\begin{figure}
    \centering
    \includegraphics[width=0.25\textwidth]{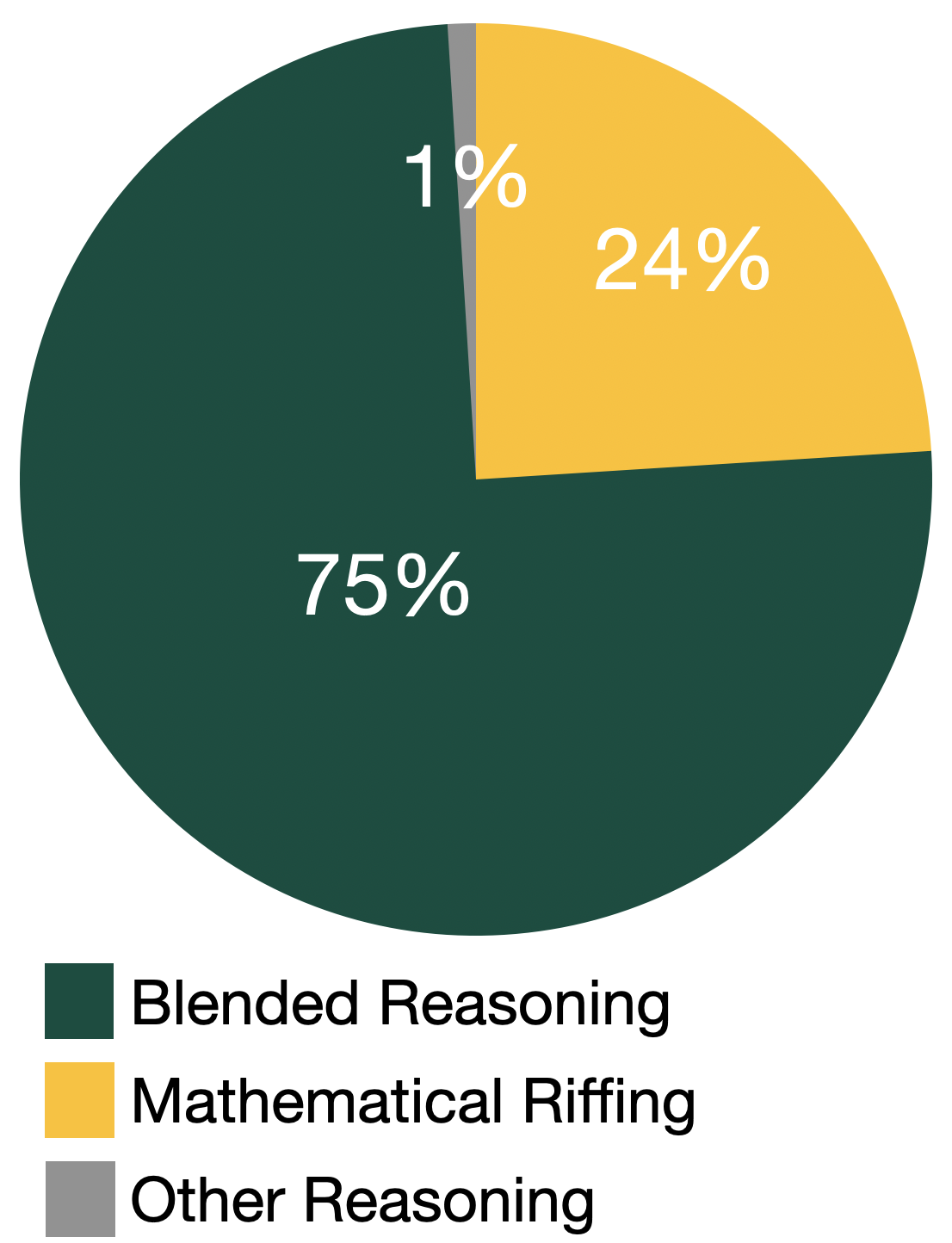}
    \caption{Percentage of time spent reasoning in a blended way, which we measure as Quantity and Task Context coded together, abstractly, which we call mathematical riffing and is measured as Quantity coded without Task Context, and other reasoning, which is measured as when Quantity is coded without Task Context, but the transcript reveals that the participants are not riffing.}
    \label{fig:sfPieAllTogether}
\end{figure}

These kinds of statements were consistent throughout the data; we found that about 3/4 of the time, Quantity codes were assigned for the same utterances as Task Context codes (Fig.~\ref{fig:sfPieAllTogether}).  We interpret this to mean that, for experts engaged in graphing tasks, the majority of their reasoning about symbolic representations is grounded in the physical context of the task they are working on.

We also observed that when the experts were making sense of quantities, the mechanism for how the quantities might affect one another was an essential tool \cite{Russ2008RecognizingScience, Russ2009MakingThinking}. For example, during the Intensity task, many experts reasoned about the physical interaction between the light and the liquid to help develop their model. Several experts noted that they couldn't imagine a glass of water changing the intensity in a measurable way:
\begin{quote}
``I don't know, it's hard to imagine a laser light being that much affected by water, but I guess so. So as soon as you went infinitely far down into a glass, there would be so much light scattered before reaching the bottom that there would be no light.''
\end{quote}
\noindent
Many used the ocean as an example, preferring to think about a body of water large enough that they could imagine the light diminishing. Of those physics experts who convinced themselves the light may be affected measurably, most then wrestled with whether a linear or exponential function would be more appropriate. This debate was either because they were considering the attenuation of the light due to the liquid, or because of self-described inexperience with the context.

Experts were also observed to use reasoning about the physical mechanism to validate their models. For example, Gravitation prompted many physics experts to invoke braking as a mechanism to explain a conflict between their assumptions and the animation. To begin the task, physics experts uniformly produced an already known model of potential energy as a function of distance from the source (either graphically or symbolically), and then used it to create new graphs of potential energy as a function of total distance traveled. However, they had expectations of how kinetic energy and potential energy would change together based on a constant total energy model.
One faculty member stated,
\begin{quote}
``It's going to speed up as it gets zooming into this thing. So I'm going to swing it down here and swing it up there because there's more kinetic energy down here [referring to the points closest to the planet Rigamarole].''
\end{quote}
\noindent
To make sense of this with the animation, which shows a spaceship moving at constant speed, several experts described braking as a mechanism to explain the motion:
\begin{quote}
``There's probably some energy source the shuttle is using when it's flying away, is my guess. And when it's braking. 'Cause I don't know how else it will take that path.''
\end{quote}

These examples illustrate how the physics experts were engaged not just in the technical definition of the quantity---its unit, sign and physical meaning---but also the \emph{physical implication} when it changes. The quantities bring to mind both particular functions or contexts and a sense of physicality; the physics experts expressed ideas about what quantities made sense to change and what mechanism would cause the change \footnote{For more information about how the physics experts reasoned about rates of change and functions in particular, please see our accompanying publication \cite{Zimmerman2023ExpertTasks}.}.

\subsection{Mathematical Riffing}
When transcript portions are categorized as Quantity but not categorized as Task Context,  we typically see physics experts setting aside the context of the task to reason algebraically. However, the algebraic reasoning we observed is not ``purely mathematical,'' that is, devoid of contextual or physical meaning. For example, one expert working on Drone stated:
\begin{quote}
    ``I think all I know is the\ldots $x$ is going to be some $v_0 t$\ldots And $v_0$ is related to $\theta_0$. So let's write $x$ of $t$.''
\end{quote}
This expert makes a series of algebraic moves based on their assumption that the problem can be modeled with a constant horizontal velocity. Their algebraic reasoning carries with it implicit physical meaning---$v_0$ represents velocity, and is introduced as an already established fact (despite this being its first occurrence in the interview). This quote is representative of what we observed across the expert pool when we reexamined sections of transcript categorized as Quantity and \emph{not} Task Context: the reasoning seemed to be anchored in already familiar physical models. These moments are outlined in Figure~\ref{fig:ganttCharts}. The timeline charts illustrate that physics experts are likely to engage with physical models before, after, and often during these periods of mathematical reasoning (as shown by the prevalence of green squares, indicating the category Models, near outlined sections of the timelines).

We name this behavior ``mathematical riffing'': the physics experts are playing around with familiar models to generate a new expression. We use this language as an analogy to the practice of riffing by jazz musicians. To the untrained ear, jazz improvisation may appear to be random. However, playing improvisational jazz is deeply informed by the musician's experience and knowledge. Phrases such as ``What happens if\ldots'' or ``Let's see what we can do with this\ldots'' were strong indicators that an expert was engaged in mathematical riffing. During this process, it was common for physics experts to refer back regularly to the task to check that what they were doing made sense---in the timeline charts, this is seen by the Task Context code category appearing in between, or at the ends of, outlined sections. These checks were used to make sense of the model they were generating and to confirm, or reject, an emerging model as productive towards solving the task. To exemplify mathematical riffing, we provide a transcript excerpt from an interview with one of the faculty members whom we call Ernest.

\begin{figure*}
    \centering
    \includegraphics[width=0.8\textwidth]{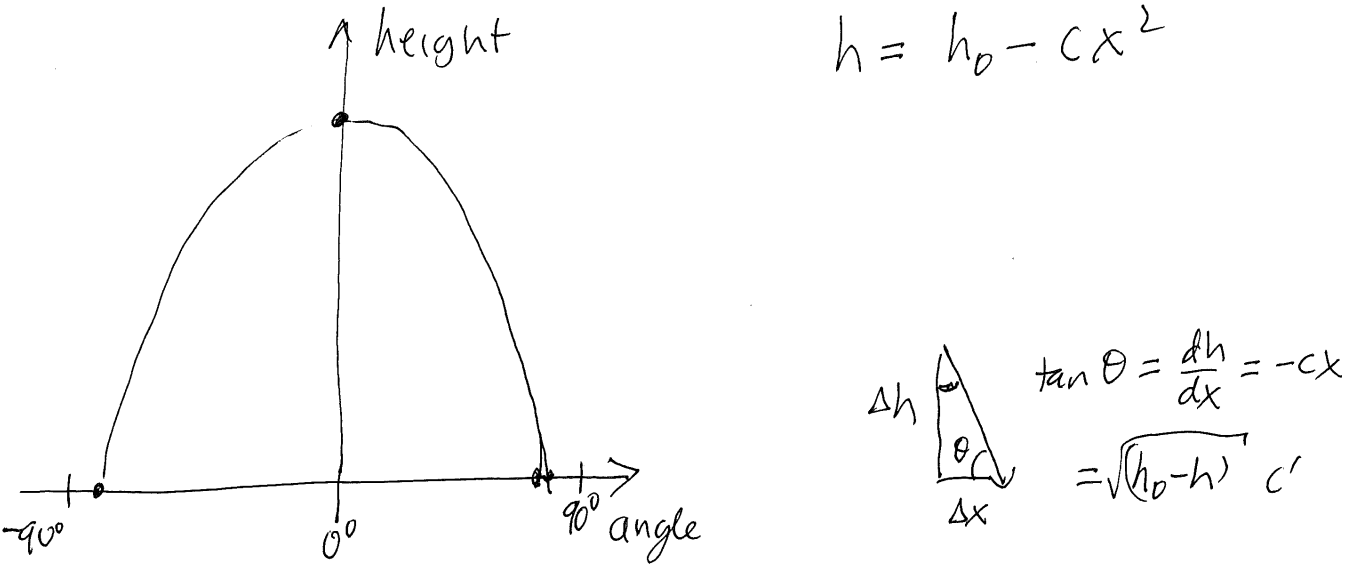}
    \caption{Ernest's written work for the Drone item.}
    \label{fig:ernestDrone}
\end{figure*}

While working on the Drone task, Ernest recognized that the parabolic shape might mean the motion of the drone could be modeled using a quadratic function. He stated:
\begin{quote}
    ``So [the height of the drone] is going to be minus some constant times $x$ squared, which we can work out.

    ``This is just assuming it's a quadratic, actually, I don't know that for sure. 

    ``Well, we said no air resistance, so we do know that for sure.''
\end{quote}
Ernest wrote $h = h_o - cx^2$ on his paper to define the height, $h$, by the horizontal position of the drone, $x$, and then checked his expression against the context of the task to make sure it made sense (see Fig.~\ref{fig:ernestDrone}). The reference to the absence of air resistance suggests the model Ernest is using may be linked to 
a model of constant acceleration in his mind, but he did not state so explicitly. Later, he worked on defining the angle, $\theta$:
\begin{quote}
    ``The tangent of this angle is going to be given by $dh$ by $dx$. 

    ``Okay, which is minus $c$ times $x$.

    ``And that's in some complicated relation to $h$, which is given by $h_0$ minus $h$, square root, thereof and some factor, which I could work out. So, okay, so this is some messy formula, I'm not going to come up with this very well.

    ``So we'll just draw a smooth curve, which goes between these two, and we'll call it a day at that point, okay.''
\end{quote}
Ernest determined a symbolic representation for both $h$ and $\theta$, and attempted to find an expression of $h$ in terms of $\theta$. However, he quickly changed tack when he realized the expression would be complicated and possibly unhelpful towards completing the task. During this time, it was clear that the symbols held physical meaning for Ernest, even when he was disconnected from the task itself.

\begin{figure}
    \centering
    \includegraphics[width=0.45\textwidth]{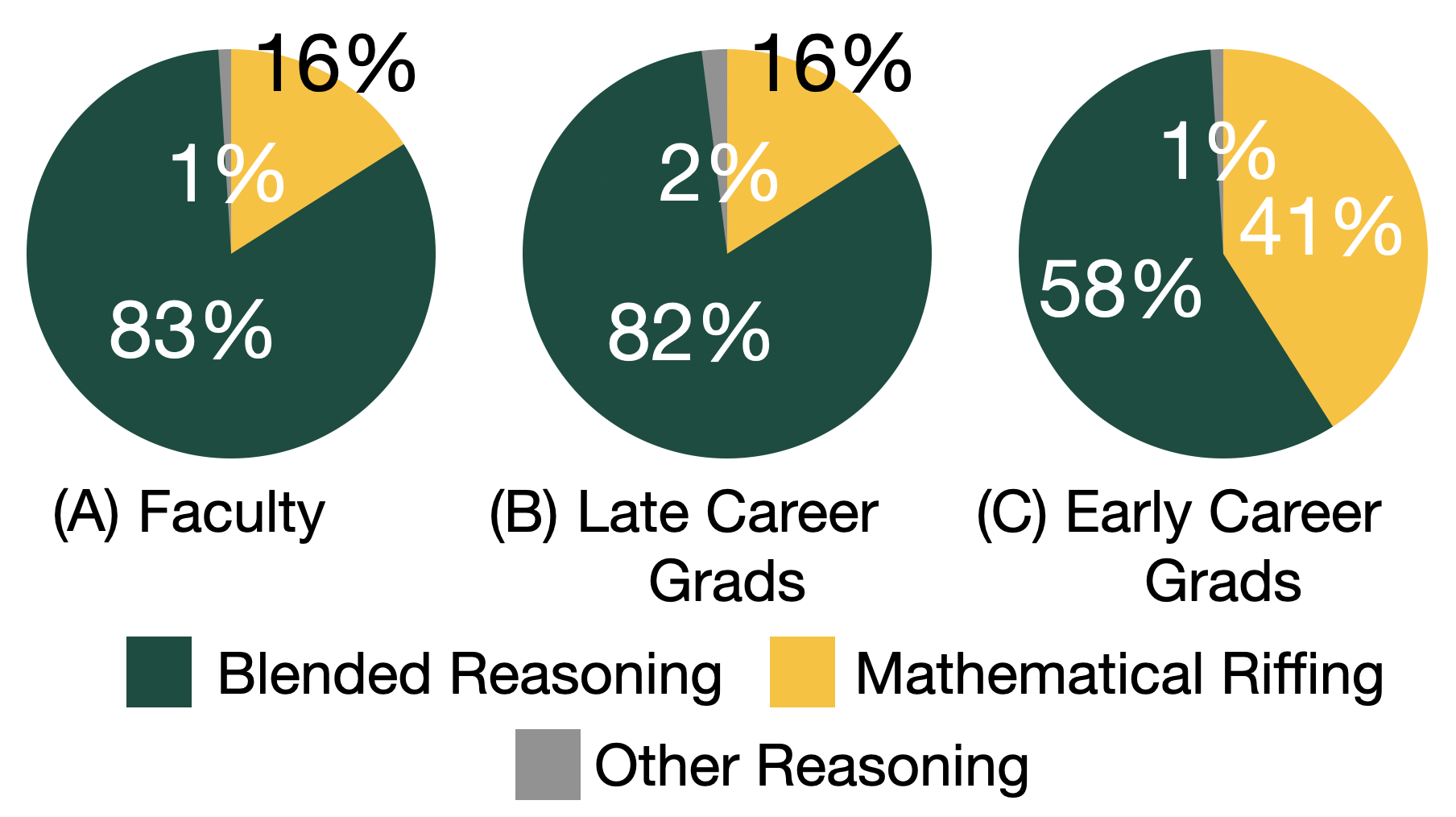}
    \caption{Percentage of time spent reasoning: in a blended way, as measured by Task Context and Quantity coded together; mathematically riffing, as measured by when Quantity is coded without Task Context; and Other, as measured by when Quantity is coded without Task Context but the transcript reveals that the reasoning is not fully aligned with our definition of mathematical riffing. Percentages were calculated as percent time on task for each participant, and averaged by population to produce the charts shown.}
    \label{fig:sfPieByPop}
\end{figure}

The majority of the physics experts engaged in mathematical riffing at some time during the interview, suggesting it is a useful tool and a mark of expertise. We observed that productive riffing is characterized by adjusting established physical models to describe a novel context. Equally important to productive riffing is the flexibility to change tack quickly when a model does not appear to work well for a given task. 

We also observed unproductive riffing, which is characterized by continuing to persevere with a model that does not get the reasoner closer to a solution. As an example, we provide a short excerpt from an early career graduate student, Peyton. While working on Drone, Peyton spent a long period of time attempting to derive an analytical expression:
\begin{quote}
    ``\ldots So right now I have $y$ is equal to $x \tan{\theta}$, which is equal to $x_0$ plus $v_x t \tan{\theta}$. I don't want to eliminate $\theta$, but I’d like to eliminate time\ldots
    
    ``If I put $y$ over $\tan{\theta}$ into here\ldots  then\ldots have something that’s quadratic in both. I don’t really like that\ldots''
\end{quote}
Peyton continued in this manner for a large percentage of their time on task, as shown by the dashed boxes in Figure~\ref{fig:ganttCharts}. Throughout their time reasoning algebraically, they seemed unconvinced that they were on the correct path, regularly making statements like ``This looks messy\ldots'' and ``That must be wrong.'' Despite their concern, they continued working with the same model for the majority of their time spent solving Drone.

We observed a difference in prevalence of mathematical riffing across populations. Faculty and late career graduate students were less likely to spend much time riffing, as shown in Figure~\ref{fig:sfPieByPop}, which we attribute to the fact that they were quick to abandon unproductive models. The early career graduate students percentages are dominated by Peyton, who engaged with mathematical riffing far more than their peers and often in unproductive ways. However, even when we removed Peyton from the data set, a trend is apparent: the average percent time on task riffing for the other four early career graduate students was 30\%. A full timeline chart of each participant can be found in the Appendix for comparison.

\section{Discussion}
\label{sec:discuss}
In this study we focus on how physics experts engaged with mathematical and physical reasoning during a series of graphing tasks. The tasks in these interviews were designed to probe graphical reasoning, and therefore did not prompt participants to generate symbolic expressions. However, many experts spontaneously chose to develop symbolic representations to help themselves reason.  We found that experts engaged in physical sensemaking whether reasoning symbolically or not; physics experts continually maintained a connection to the physical world throughout their interviews. There was little to no evidence that the physics experts were ever reasoning in a purely mathematical space, that is, one in which symbols hold no physical meaning. This finding suggests that  mathematical reasoning in the contexts of the chosen tasks almost always contains some reference to the real world for physics experts. 

It was common for the physics experts to reason with specific functions to help them reason through the graphing tasks, which may feel to them like they are doing ``pure mathematics.'' However, we found there was always physical reasoning embedded in their mathematical reasoning---the variables carried physical meaning and experts had expectations about how they would behave as another quantity changed. These periods of algebraic reasoning, what we are calling mathematical riffing, are characterized by physics experts manipulating already familiar physical models, seeing what happens, and either noticing when the model is not fruitful for them and abandoning the line of reasoning, or building on the model. Essential to mathematical riffing is that the expert bases their reasoning on expected common functions in physics and on familiar physics models. The expert manipulates these models in a way that may appear random or inexpert-like to a novice, but is informed by an understanding of what makes sense for the context. This finding suggests that physics experts associate physical meaning with algebraic variables in ways that mathematicians typically don't, and that it is an essential part of mathematical thinking in physics.

We also found that the physical mechanism by which quantities changed was integral to how physics experts reasoned about the meaning of their representations. Research has shown that using the physical mechanism of a system to make sense of why a quantity may change is an important aspect of conceptual physics reasoning \cite{Hammer2003TappingPhysics, Russ2008RecognizingScience, Richards2013HowTopics, Jones2015CognitiveExpertsb}.  Euler, R\r{a}dahl, and Gregorcic describe how mechanistic reasoning is an important part of students developing explanatory models; research in physics education has demonstrated the value of explanatory models in modeling instruction \cite{Etkina2006TheInstruction, Euler2019EmbodimentLook}. This study provides further evidence that mechanistic reasoning also informs the development of mathematical models by physics experts.

Finally, we observed differences in the effectiveness of mathematical riffing across our expert populations. Figure~\ref{fig:sfPieByPop} shows that the early career graduate students in our sample spent more time mathematically riffing than more experienced experts, because they were less likely to reject an unproductive line of reasoning quickly. This emergent pattern is aligned with Bing and Redish's description of a journeyman: ``that level where students have developed sufficient skills that they can no longer be considered novices but where they have not yet had sufficient experience with sophisticated problem solving (and research) to be considered experts'' \cite{Bing2012EpistemicTransition}. Bing and Redish compared novice and journeymen reasoning, and showed that journeymen more quickly recognized their approach was unproductive and were more likely to change tack than novices. Our data represent another section of the spectrum, from journeyman to expert. Together, these results demonstrate that rejecting unproductive models is a hallmark of expertise, and that this skill may be slow to develop across many years of physics experience. 

Given the small sample size and the potential for embarrassment while  problem solving in an interview, we emphasize that further research is needed here before we strongly claim  that there is an ``early-career'' effect. We cannot ignore the potential influence of imposter syndrome---the participants were aware the study was focused on expert reasoning, and this may have had an impact on how they felt during the interview. 

In summary, we interpret our findings that, in the context of challenging, novel, classical physics graphing tasks, physics experts continually maintain a tie to the real world, even when reasoning mathematically. This finding is aligned with those of Czocher and Serbin and Wawro \cite{Czocher2016IntroducingThinking, Serbin2022TheProblems}. Despite the differences in methodology and context---both Czocher and Serbin and Wawro examined tasks that prompted symbolic answers, interviewed undergraduate students in engineering and physics respectively, and used \textit{a priori} coding schemes based on established modeling cycles---the results are strikingly similar. These complementary findings do not support the prevalence of frameworks that separate mathematical reasoning and reasoning about the real world; in the contexts encountered by the vast majority of students who take physics courses, rather, the findings support descriptions of modeling that incorporate a continuous interaction between mathematical reasoning and physics reasoning. Future research might consider revising theoretical frameworks of modeling to support expert behaviors of using physical models as a basis for reasoning mathematically.

We encourage instructors to reflect on their thinking in the context of modeling by noticing how and when physical ideas are incorporated, often implicitly, when ``doing math'' in physics contexts. Even considerations such as why particular arithmetic operations make sense in a model often have physical foundations (i.e., summation in conservation laws, subtraction and division in rates of change). Acknowledging this subtle expert knowledge might help students recognize what is familiar, and what is new about the modeling that they are learning to do in physics. 

\section{Conclusion}
\label{sec:conclusion}
In this paper, we present empirical evidence of the emergent reasoning of physics experts  while they develop graphical models.  While there is a spectrum of reasoning from more math-like to more physics-like, there is scant evidence in our data of purely mathematical reasoning. The majority of the time, physics experts' reasoning involved interwoven mathematical and physical reasoning. When experts reasoned mathematically without considering the context of the task at hand, they engaged in mathematical riffing---using abstract physical models as a starting point, adjusting the mathematics to work towards a novel expression, and quickly rejecting models that were not productive. Considering whether models were consistent with physical mechanisms played an important role in their decision-making.  

This study was limited by the context of the tasks, which were all prompting for graphical representations (although several experts spent a good fraction of their time generating and engaging in symbolic representations) of some aspect of constant speed motion in classical physics contexts. The narrow scope of the contexts correspondingly limits the generalizability of the results; we do not claim that our findings characterize physics reasoning in every context.

Additional research can help span the space of expert mathematical reasoning in physics.  For example, we can imagine enriching these findings by replicating this study in more abstract contexts, such as quantum or statistical mechanics. While it was difficult to generate prompts that are actually novel for experts, our findings suggest that the process of examining experts generating models is potentially fruitful for future research.

Basing instruction on an assumption a mathematical model is isolated from its real world, or physical, context is not consistent with the way that physics experts appear to model. Instruction that is implicitly based on mathematical procedures isolated from physical meaning might serve to increase the gap between students and their instructions, who largely don't reason this way. We encourage instructors to reflect on their own experiences of ``doing math'' in physics, and to be aware of implicit assumptions about deeper physical meanings that are nearly always embedded in the mathematical reasoning of introductory physics. In addition, instructors, specifically when rejecting a model, may find it productive to consider ways to make their decision-making process explicit to the students in their physics courses. In general, instructional interventions that help bridge gaps between what experts may consider foundational mathematical reasoning in physics and what novices are prepared for in their prerequisite courses will enrich the learning of physics. Our findings can help the ongoing and future efforts to close these gaps. 

\section{Acknowledgements}
We would like to thank the faculty and students at the University of Washington who have made this work possible---their participation and support has been integral to this project. This work was partially supported by the National Science Foundation under grants No. DUE-1832836, DUE- 1832880, DUE-1833050, DGE-1762114, DUE-2214765, and DUE-2214283. Some of the work described in this paper was performed while the second author held an NRC Research Associateship award at Air Force Research Laboratory.

\bibliography{references}

\clearpage
\onecolumngrid
\appendix*

\section{}
\label{sec:app}
\onecolumngrid
\begin{figure*}[h!]
    \centering
    \includegraphics[width=0.95\textwidth]{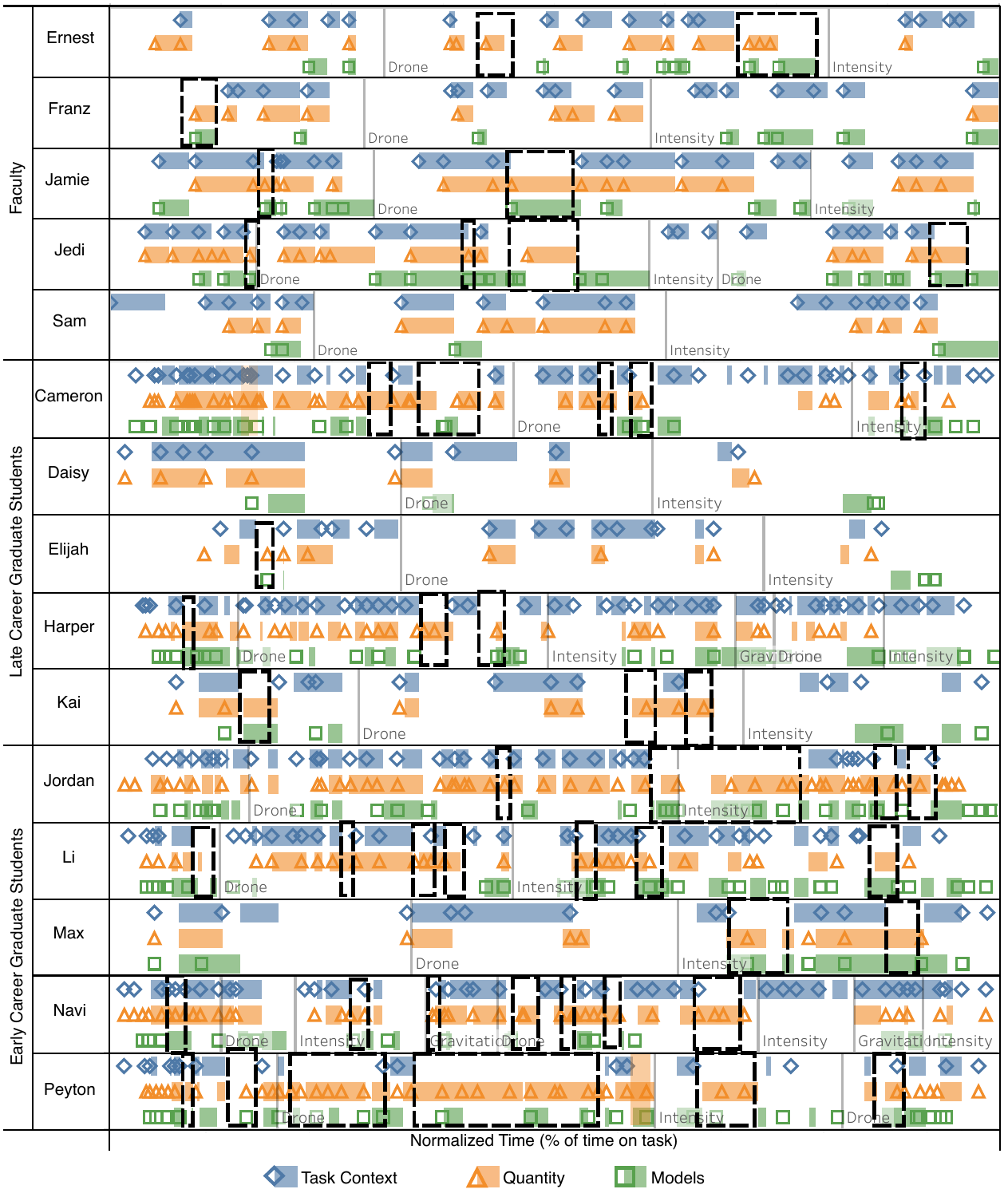}
    \caption{A visual representation of when participant statements were assigned the ``Quantity,'' ``Physical World,'' and ``Models'' code categories. The length of the bars corresponds to the time the participant was engaged in that line of reasoning. The horizontal axis represents time, normalized to the length of each interview. Therefore, it can be thought of as a percentage of their total time on task. The highlighted regions are where ``Quantity'' appears without ``Physical World.''}
    \label{fig:SFganttcharts}
\end{figure*}

\end{document}